\documentclass{article}

\usepackage{microtype}
\usepackage{graphicx}
\usepackage{subfigure}
\usepackage{booktabs} 

\usepackage{hyperref}

\usepackage{amsmath}
\usepackage{multirow}
\usepackage{makecell}
\usepackage{float}

\usepackage[accepted]{icml2019}

\icmltitlerunning{An Online Topic Modeling Framework with Topics Automatically Labeled}

\begin{document}

\twocolumn[
\icmltitle{An Online Topic Modeling Framework with\\ Topics Automatically Labeled}




\begin{icmlauthorlist}
\icmlauthor{Fenglei Jin}{CUHK}
\icmlauthor{Cuiyun Gao}{CUHK}
\icmlauthor{Michael R. Lyu}{CUHK}
\end{icmlauthorlist}

\icmlaffiliation{CUHK}{Department of Computer Science and Engineering, The Chinese University of Hong Kong, Hong Kong, China}

\icmlcorrespondingauthor{Fenglei Jin}{fengleijin@gmail.com}
\icmlcorrespondingauthor{Cuiyun Gao}{cygao@cse.cuhk.edu.hk}

\icmlkeywords{Deep Learning, Topic Extraction}

\vskip 0.3in
]



\printAffiliationsAndNotice{} 

\begin{abstract}
In this paper, we propose a novel online topic tracking framework, named IEDL, for tracking the topic changes related to deep learning techniques on Stack Exchange and automatically interpreting each identified topic. The proposed framework combines the prior topic distributions in a time window during inferring the topics in current time slice, and introduces a new ranking scheme to select most representative phrases and sentences for the inferred topics in each time slice. Experiments on 7,076 Stack Exchange posts show the effectiveness of IEDL in tracking topic changes and labeling topics.
\end{abstract}

\section{Introduction}
\label{intro}

Recent advances in deep learning promote the innovation of many intelligent systems and applications such as autonomous driving and image recognition. Tracking the changes of focus for deep learning engineers and researchers is helpful to identify current emerging deep learning-related topics. In this work, we choose Stack Exchange to collect experimental dataset due to its popularity among the developers and researchers~\cite{DBLP:conf/kbse/HuangXXLW18}.

Previous topic tracking approaches~\cite{DBLP:conf/icml/BleiL06,DBLP:conf/icdm/AlSumaitBD08,DBLP:journals/tist/HeLGW13} are mainly based on Latent Dirichlet Allocation (LDA) \cite{DBLP:journals/jmlr/BleiNJ03}. For example, the work~\cite{DBLP:conf/icdm/AlSumaitBD08} proposes an On-line Latent Dirichlet Allocation (OLDA) model to capture the evolution of topics, where only the topic distribution of documents in the prior one time slice is considered for inferring the topics in current time slice. In~\cite{DBLP:journals/tist/HeLGW13}, the authors focus on modeling sentiment and topic changes synchronously, and the topics in all the prior time slices are involved during inferring the current topic distribution. In~\cite{DBLP:journals/corr/abs-1805-00457}, the proposed approach also selects sentiment words for each topic based on Dynamic Topic Model (DTM)~\cite{DBLP:conf/icml/BleiL06}. 

Inspired by the recent work~\cite{DBLP:conf/icse/GaoZLK18}, where an adaptively online latent Dirichlet allocation approach, named IDEA, is introduced to track user opinions in user feedback, and outperforms the OLDA approach~\cite{DBLP:conf/icdm/AlSumaitBD08}, we propose a new framework \textbf{IEDL} for \textbf{I}dentifying \textbf{E}merging \textbf{D}eep \textbf{L}earning-related topics. The difference between IDEA and our approach lies in the combination styles of the prior topic distributions. In IDEA, the similarities between the topics in previous time slices and those in the previous one time slice are taken into account for inferring the topics in current time slice, while we introduce an exponential decay function in a time window. Besides, we propose a novel topic labeling approach based on the unique characteristics of Stack Exchange posts.

The experimental results on 7,076 Stack Exchange posts verify the effectiveness of IEDL in detecting topic changes and topic labeling.

The contributions of our paper are elaborated as below.
\begin{itemize}
\item We propose a framework called IEDL to automatically track topic changes and identify emerging topics from deep learning-related posts in Q\&A forum effectively.
\item We propose a novel topic interpretation method, which improve the topic coherence dramatically.
\item We visualize the variations of the captured (emerging) topics along with time slices, with the emerging ones highlighted.
\end{itemize}

\section{Methodology of IEDL}
IEDL mainly contains two parts: Emerging topic detection and automatic topic interpretation. 

\subsection{Emerging Topic Detection}
In this section, we aim to detect the emerging topics of current time slice by considering the topics in previous time slices. We first introduce how we use online topic modeling to capture the topic evolutions with time going by. Then we present how we discover the emerging topics
(anomaly topics).

\subsubsection{Online Topic Modeling}
\begin{figure}[ht]
\begin{center}
\centerline{\includegraphics[width=0.85\columnwidth]{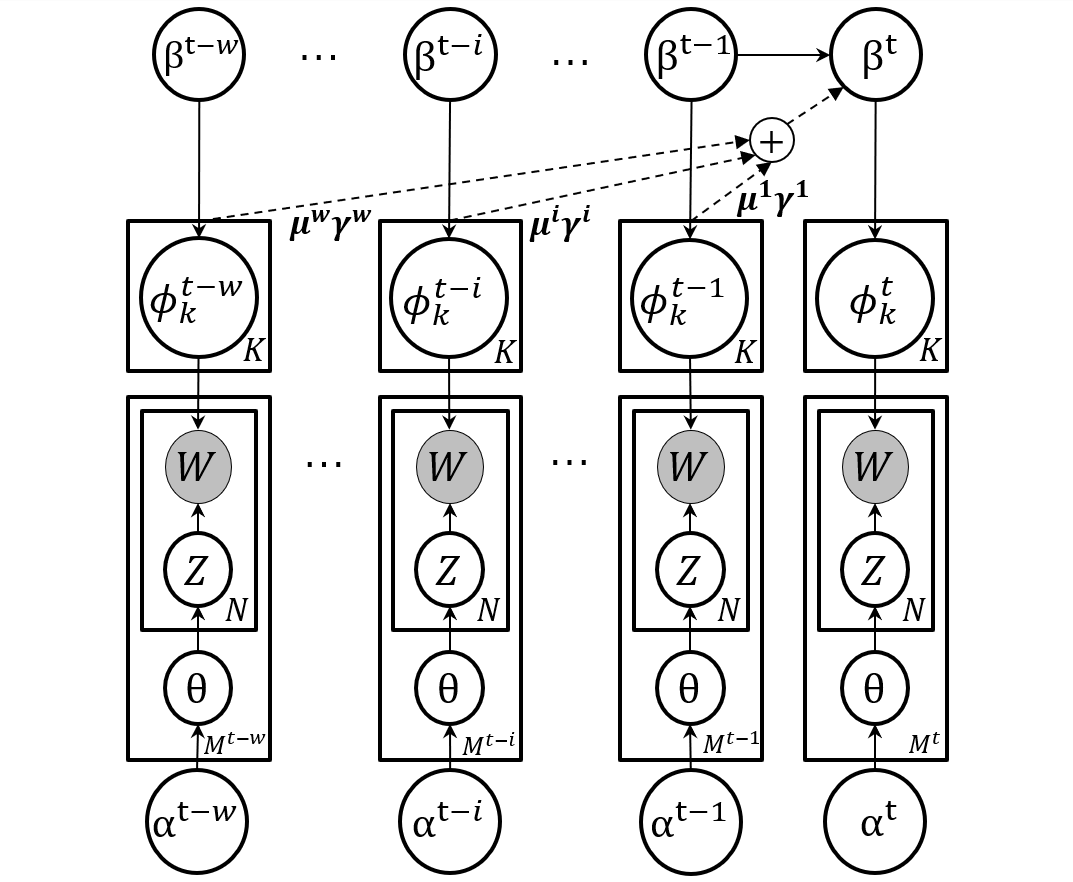}}
\caption{Our online topic modeling approach.}
\label{aoldap}
\end{center}
\end{figure}

The preprocessed posts are divided by time, denoted as $M=\{M^1, M^1,...,M^t...\}$ (where $t$ indicates $t$-th month in our experiment), and each post is treated as one document. The prior distributions over document-topic ($\alpha$) and topic-word distributions ($\beta$) are defined initially. $K$ represents the number of the topics, while $\phi^t_k$ is the probability distribution vector for the $k$-th topic over all the input posts. We also introduce a predefined parameter - window size $w$, which refers to the number of previous time slices to be considered for analyzing the topic distributions of the current time slice. The overview of the model is shown in Figure\ref{aoldap}.

We adaptively integrate the topic distributions of the previous $w$ time slices, denoted as $\{\phi^{t-1},...,\phi^{t-i},...,\phi^{t-w}\}$, for generating the prior $\beta^t$ of the $t$-th time slice. Since the popularity of a topic always lasts for a time period, compared to IDEA~\cite{DBLP:conf/icse/GaoZLK18} which only considers the similarity between topic distributions, we think the topics discussed last month are more related to current topics compared to those mentioned several months before. Therefore, an exponential decay factor $\mu$ is added, multiplying the similarity between topics $\gamma$ to determine the influence. And now the adaptive integration refers to sum of the topic distributions of different time slices with different weights $\gamma^i$ and $\mu^i$:
\begin{equation}
    \beta^t_k = \sum^w_{i=1}\mu^i\gamma^{i}_k\phi^{t-i}_k
\end{equation}
where $i$ denotes the $i$-th previous time slice $(1 \leq i \leq w)$. We denote the weight $\gamma^i$ as the similarity of topic distributions between the $(t$$-$$i)$-th time slice and the $(t$$-$$1)$-th time slice, which is calculated by the softmax function:
\begin{equation}
    \gamma^i_k = \frac{\exp(\phi^{t-i}_k \cdot \beta^{t-1}_k)}{\sum^w_{j=1}\phi^{t-j}_k \cdot \beta^{t-1}_k}
\end{equation}
 where the dot product $(\phi^{t-i}_k \cdot \beta^{t-1}_k)$ computes the similarity between the topic distribution $\phi^{t-i}_k$ and the prior of the $(t$$-$$1)$-th time slice $\beta^{t-1}_k$. $\mu^i$ is calculated by a simple exponential decay function:
\begin{equation}
    \mu^i = \exp(-\lambda i)
\end{equation}
where $i$ means the $i$-th time slice before the current, and $\lambda$ is a predefined exponential decay coefficient.

\subsubsection{Anomaly Discovery}
Based on the topic distribution captured by online topic model, we regard anomaly topics, which present obvious distinctions compared to those of the previous time slices, as emerging topics. To calculate the distinction of the $k$-th topics between two successive time slices, we implement the classic Jensen-Shannon (JS) divergence\footnote{\url{https://en.wikipedia.org/wiki/Jensen\%E2\%80\%93Shannon\_divergence}}. If we take $\phi^t_k$ and $\phi^{t-1}_k$ as an example:
\begin{equation}
\begin{split}
    D_{JS}(\phi^{t}_k||\phi^{t-1}_k) = \frac{1}{2}&D_{KL}(\phi^{t}_k||M) +\\ &\frac{1}{2}D_{KL}(\phi^{t-1}_k||M)
\end{split}
\end{equation}
where $M=\frac{1}{2}(\phi^{t}_k+\phi^{t-1}_k)$.And the Kullback-Leibler (KL) divergence $D_{KL}$ is utilized to measure the difference from one probability distribution $P$ to another $Q$:
\begin{equation}
    D_{KL}(P||Q) = \sum_i P(i)\log\frac{P(i)}{Q(i)}
\end{equation}
where $P(i)$ is the $i$-th item in P. The higher the JS divergence is, the larger distinction the two topic distributions have. To find anomaly topics, we set a threshold by leveraging a typical outlier detection method \cite{DBLP:journals/widm/RousseeuwH11}. For each time slice, the topics with divergences higher than the threshold are regarded as anomaly topics.

\subsection{Automatic Topic Interpretation}
The dataset we use are questions asked in Stack Exchange, which have two significant attributes ``votes" and ``views". Users can manually click the ``like" button or the ``dislike" button to show their preference, while high ``votes" represents this is a valuable question. And ``views" shows how many users or tourists have visited this page, which refers to the popularity of this post. To make good use of these two attributes, we develop a novel method for deep learning related posts interpretation. We denote it as Quality Score:
\begin{equation}
\begin{split}
    SC&ORE_{qua}(l) =\\ &\exp(-\frac{1}{\ln(v_l+1)\ln(r_l+1)}- \eta\cdot\frac{1}{\ln(h_l+1)})
\end{split}
\end{equation}
where $l$ is the post, and $v_l$, $r_l$, $h_l$ are the votes, views, and length of the post respectively. The attributes are adding $1$ in case of they are $0$. If a post has both high votes and views, it is more likely to be a good post. Also, length slightly influence the score by the predefined factor $\eta$, since long posts may contain more information. Therefore, the motivation of this Quality Score is to select questions with both high votes and views with longer length.

\section{Experiment and Result}
In this section, we introduce how we preprocess the dataset, and the performance of our IEDL model measured by topic distribution classification precision and topic coherence.

\subsection{Data Analysis}
The 7,076 deep learning-related posts we used are publicly released by Stack Exchange\footnote{\url{https://archive.org/download/stackexchange}}. To evaluate the topics inferred by our proposed topic model, we also manually labeled 507 posts into six categories for classification: Image, NLP, Game-ai, Self-driving, Programming-languages, and Reinforcement-learning. The labels are determined based on the tags provided by Stack Exchange and to maximize their distinguishability. 

\subsubsection{Word Formatting}
We first convert all words into lowercase, and then perform lemmatization to change each word into its original form. We then replace some segments with general symbols, like converting websites to ``$<$url$>$'' and so on.

\subsubsection{Phrase Extraction}
Since some words have specific meanings only in phrases and we need them to interpreting topics, phrases (mainly referring to two
consecutive words in our paper, and the words in each phrase are connected with ``\_") are  extracted in the preprocessing step and trained along with all the other words. We want the topic labels in phrases to be meaningful and comprehensible, therefore, a typical phrase extraction method based on PMI (Pointwise Mutual Information)\footnote{\url{https://en.wikipedia.org/wiki/Pointwise\_mutual\_information}}, which is effective in identifying meaningful phrases based on co-occurrence frequencies, is used:
\begin{equation}
    PMI(w_i,w_j) = \log\frac{p(w_{i}w_{j})}{p(w_i)p(w_j)}
\end{equation}
where $p(w_{i}w_{j})$ refers to the co-occurrence probability of the phrase $w_{i}w_{j}$ and $p(w_i)$ and $p(w_j)$ indicates the probability of the word $w_i$ and $w_j$ in the whole post documents. High PMI values indicate that it is more likely for the combination of the two words to be a meaningful phrase. We experimentally set a threshold for PMI, and  phrases with higher PMIs are extracted.

\subsubsection{Filtering}
This step aims to eliminate non-meaningful words, such as emotional words (e.g., ``nice" and ``bad"), abbreviations (e.g., ``btw"), and useless words (e.g., ``something"). We use the predefined stop words provided by NLTK\footnote{\url{http://www.nltk.org/}}, and all words in the stop word list are filtered out. Finally, all remaining words and extracted phrases are fed into the model.

\subsection{Classification Accuracy}
To test the quality of the extracted topics, we use the topic distribution of each post as features, and classify the 507 labeled posts by SVM. The results show that our proposed model outperforms the baseline model IDEA~\cite{DBLP:conf/icse/GaoZLK18} by 5\% for average precision.
\begin{table}[h]
\caption{Classification result}
\small
\label{classification}
\begin{center}
\begin{sc}
\scalebox{0.85}{\begin{tabular}{|c|c|ccc|} \hline 
Category & Model & \scriptsize{Precision} & \scriptsize{Recall} & \scriptsize{F1}\\ \hline
\multirow{2}{*}{Image} & IDEA & 0.89 & \bf 0.73 & \bf 0.80 \\ 
\cline{2-5} & IEDL & \bf 1.00 & 0.64 & 0.78 \\ \hline
\multirow{2}{*}{NLP} & IDEA & 0.68 & 0.76 & 0.72 \\ 
\cline{2-5} & IEDL & \bf 0.73 & \bf 0.94 & \bf 0.82 \\ \hline
\multirow{2}{*}{Game-ai} & IDEA & 0.83 & 0.94 & 0.88 \\ 
\cline{2-5} & IEDL & \bf 0.83 & \bf 0.97 & \bf 0.90 \\ \hline
\multirow{2}{*}{Self-driving} & IDEA & 0.94 & 0.89 & 0.91 \\
\cline{2-5} & IEDL & \bf 1.00 & \bf 0.94 & \bf 0.97 \\ \hline
\multirow{2}{*}{\makecell{Programming\\-language}} & IDEA & \bf 0.92 & 0.73 & 0.81 \\
\cline{2-5} & IEDL &  0.86 & \bf 0.86 & \bf 0.86 \\ \hline
\multirow{2}{*}{\makecell{Reinforcement\\-learning}} & IDEA & 0.86 & \bf 0.86 & \bf 0.86 \\ 
\cline{2-5} & IEDL & \bf 1.00 & 0.62 & 0.76 \\ \hline
\end{tabular}
}
\end{sc}
\end{center}
\end{table}

\subsection{Topic Coherence}
\begin{table}[h] 
\centering
\small
\caption{Topic coherence of different approaches.}
\begin{sc}
\scalebox{0.85}{\begin{tabular}{|c|c|c|c|} \hline 
OLDA & IDEA & IDEA+Quality Score & IEDL\\ \hline 
0.133 & 0.166 & 0.217 & \textbf{0.222} \\ \hline 
\end{tabular}
}
\end{sc}
\end{table}

\setcounter{figure}{2}
\begin{figure*}[ht]
\centering
\includegraphics[width=0.9\textwidth]{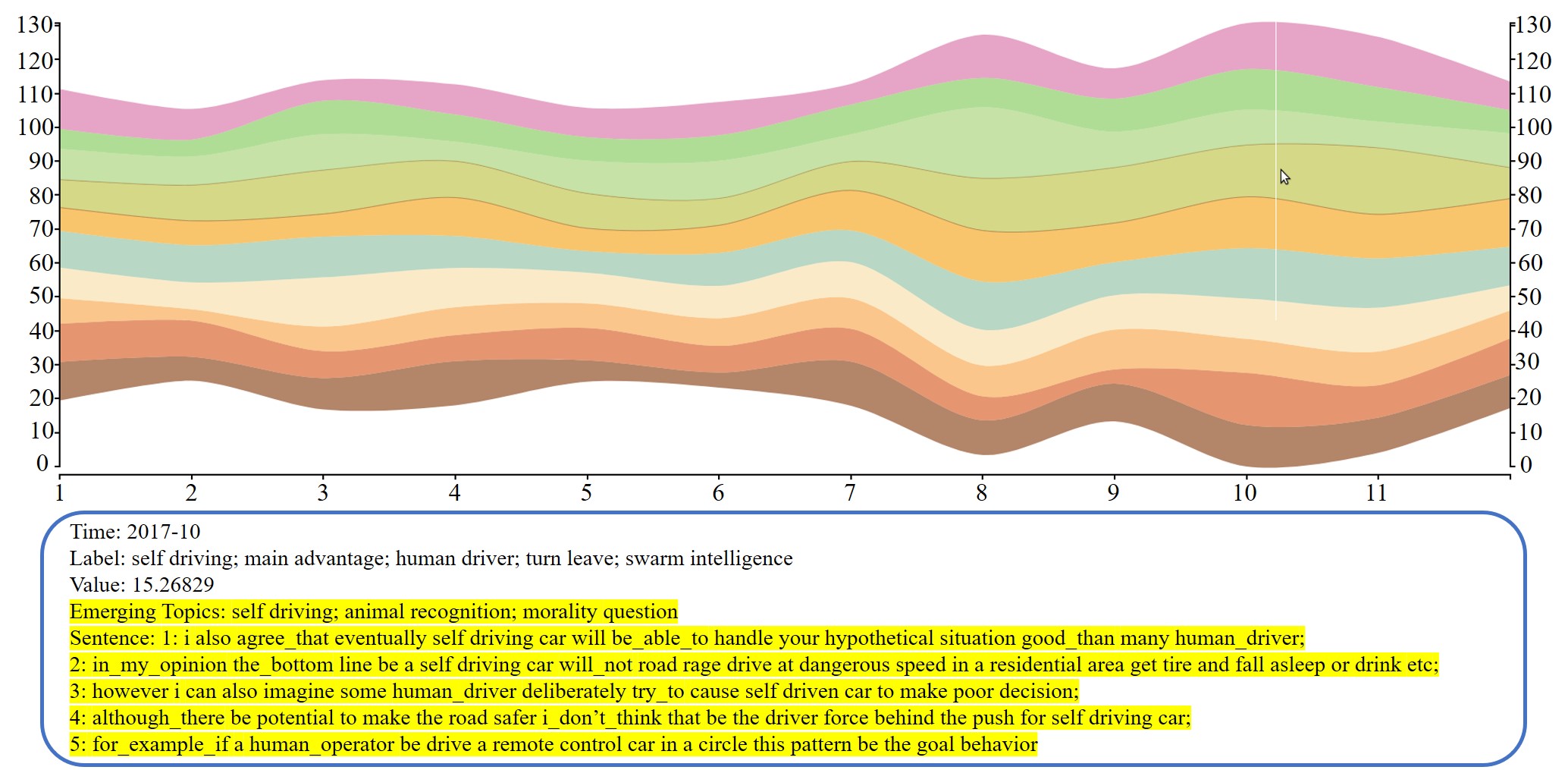}
\caption{Visualization of topic changes based on ThemeRiver~\cite{DBLP:conf/infovis/HavreHN00}. Texts highlighted in yellow are the emerging topics in the corresponding month (Oct. 2017) where the mouse is pointing at.}\label{river}
\end{figure*}

Topic coherence score~\cite{DBLP:conf/naacl/LauB16} is another way to measure the performance of models by detecting the coherence between extracted words or phrases assigned to each topic. The method we use is an extension of PMI, where $t$ refers to a topic, and $N$ is the number of words:
\begin{equation}
OC\_Auto\_PMI(t)=\sum^N_{j=2}\sum^{j-1}_{i=1}\log\frac{p(w_{i}w_{j})}{p(w_i)p(w_j)}
\end{equation}
We feed the whole 7,076 dataset into our IEDL model and compare the topic coherence score to IDEA~\cite{DBLP:conf/icse/GaoZLK18}. The result shows IEDL improves the topic coherence score by 33.7\%. The topic coherence scores with error bars are shown in Figure~\ref{TC}.

\setcounter{figure}{1}
\begin{figure}[ht]
\begin{center}
\centerline{\includegraphics[width=0.95\columnwidth]{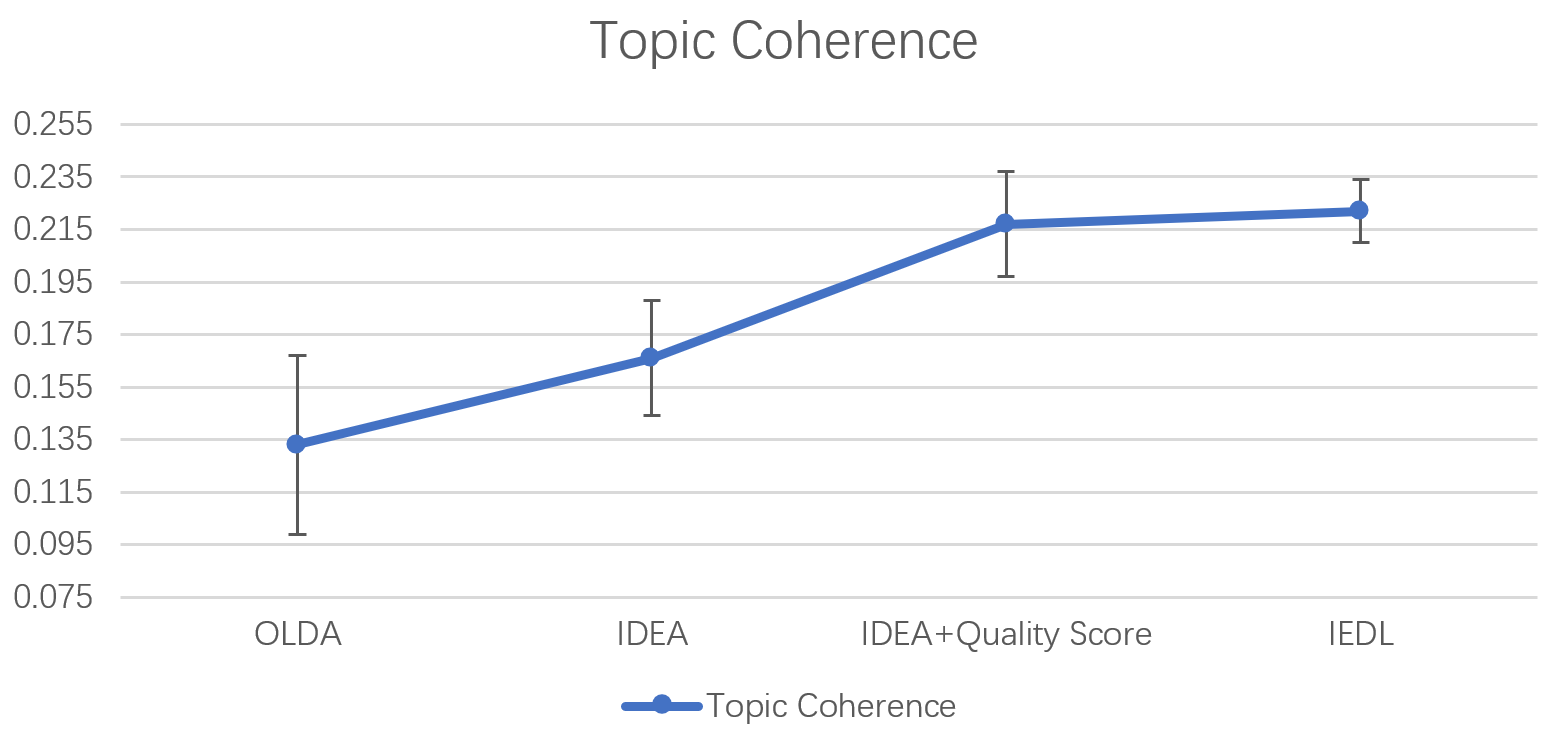}}
\caption{Topic coherence with error bars (standard error).}
\label{TC}
\end{center}
\end{figure}

To further elaborate on the coherence between topics and extracted phrases, we compared the generated phrases for topics ``NLP" and ``Image" respectively. The result shows IEDL can generate more coherent and meaningful phrases for each topic.

\begin{table}[h]
\caption{Phrases generated by IDEA and IEDL for topics ``NLP" and ``Image" respectively. Red underlined fonts highlight the phrases that are not closely related to the topic.}
\small
\label{compare}
\begin{center}
\scalebox{0.85}{\begin{tabular}{|c|c|c|c|}
  \hline
  \multicolumn{2}{|c}{NLP} & \multicolumn{2}{|c|}{Image} \\  
  \hline
   IDEA & IEDL & IDEA & IEDL \\
  \hline
  {\color{red} \underline{solution space}} & word vector &cnn model & \begin{tabular}{@{}c@{}}convolutional \\ network\end{tabular}   \\
  \begin{tabular}{@{}c@{}}information \\ science\end{tabular}  & word embedding & previous layer & pixel value  \\
  {\color{red} \underline{real environment}} & feature extraction & {\color{red} \underline{specific task}} & \begin{tabular}{@{}c@{}}capsule \\ network\end{tabular} \\ \hline
\end{tabular}
}
\end{center}
\end{table}

\section{Visualization}

In this part, we visualize the the evolution deep learning topics along with time flow for better understanding. As shown in Figure~\ref{river}, all the posts constitute one river and each branch of the river indicates one topic. By moving the mouse over one topic, one can track detailed topic changes along with time slices (months in our experiment), where the emerging issues are highlighted.

The topics with wider branches are of greater concern to developers, where the width of the $k$-th branch in the $t$-th version is defined as:
\begin{equation}
width^t_k = \sum_{a}\log Count(a) \times SCORE_{qua}(l_a)
\end{equation}
where $Count(a)$ is the count of the phrase label $a$ in the post collection of the $t$-th version, and $Score_{qua}(l_a)$ denotes the quality score of $l_a$, which is the post refers to the phrase label $a$.

We visualize topic changes from January to December 2017. As shown in Figure~\ref{river}, our IEDL finds an emerging topic about self-driving, which is not detected by IDEA~\cite{DBLP:conf/icse/GaoZLK18}. We double check the dataset and find that, compared to only one post from July to September, there are eight posts about self-driving in October (may be caused by a new release of electric semi-truck of Tesla), which further proves the effectiveness of our model in detecting emerging topics.

\section{Conclusion and Future Work}
Timely and effectively detecting deep learning topics is crucial for developers to capture the trend. We propose IEDL, a novel framework for automatically identifying emerging topics from posts in Q\&A forums. The experiment results show IEDL improves the quality of topic distribution and topic coherence greatly. In the future, we will refine IEDL to be capable of defining the topic number automatically, and utilize other information like comments, accepted answers to further improve the performance.

\bibliographystyle{icml2019}
\bibliography{example_paper}

\end{document}